\documentclass[11pt]{article}
\usepackage{graphicx}
\usepackage{amsfonts}
\usepackage{theorem}
\newtheorem{Lem}{Lemma}[section]
\newtheorem{Def}[Lem]{Definition}
\newtheorem{The}[Lem]{Theorem}

\newtheorem{Rem}[Lem]{Remark}

\newcommand{\qed}{\hbox{\rule{6pt}{6pt}}}

\begin{document}
\title{Schr\"odinger  uncertainty relation with Wigner-Yanase skew information}
\author{Shigeru Furuichi$^1$\footnote{E-mail:furuichi@chs.nihon-u.ac.jp}\\
$^1${\small Department of Computer Science and System Analysis,}\\
{\small College of Humanities and Sciences, Nihon University,}\\
{\small 3-25-40, Sakurajyousui, Setagaya-ku, Tokyo, 156-8550, Japan}}
\date{}
\maketitle
{\bf Abstract.} We shall give a new Schr\"odinger type uncertainty relation for a quantity representing a quantum uncertainty, 
introduced by S.Luo in \cite{Luo1}.
Our result improves the Heisenberg uncertainty relation shown in \cite{Luo1} for a mixed state.

\vspace{3mm}

\vspace{3mm}

\vspace{3mm}



\section{Introduction}
In quantum mechanical system, the expectation value of an observable (self-adjoint operator) $H$ 
in a quantum state (density operator) $\rho$ is expressed by $Tr[\rho H]$.
Also, the variance for a quantum state $\rho$ and an observable $H$ is defined by 
$V_{\rho}(H)\equiv Tr[\rho \left(H-Tr[\rho H]I\right)^2] = Tr[\rho H^2] -Tr[\rho H]^2$.
It is famous that we have the Heisenberg uncertainty relation \cite{Hei}:
\begin{equation}   \label{HUL}
V_{\rho}(A) V_{\rho}(B) \geq \frac{1}{4}\vert Tr[\rho[A,B]]\vert^2 
\end{equation}
for a quantum state $\rho$ and two observables $A$ and $B$.
The further strong result was given by Schr\"odinger \cite{Schr}:
\begin{equation} \label{S_UL}
V_{\rho}(A) V_{\rho}(B)-\vert Re\left\{ Cov_{\rho}(A,B) \right\} \vert^2 \geq \frac{1}{4}\vert Tr[\rho[A,B]]\vert^2,  
\end{equation}
where the covariance is defined by $Cov_{\rho}(A,B) \equiv Tr[\rho \left(A-Tr[\rho A]I\right)\left(B-Tr[\rho B]I\right)].$

On the other hand, as a degree for non-commutativity between a quantum state $\rho$ and an observable $H$, the Wigner-Yanase skew information $I_{\rho}(H)$
was defined in \cite{WY} (See Definition \ref{def1} in Section \ref{sec2}). 
It is famous that the convexity of the Wigner-Yanase-Dyson skew information 
$I_{\rho,\alpha}(H)\equiv \frac{1}{2}Tr\left[   \left( i[\rho^{\alpha},H]\right) \left( i[\rho^{1- \alpha},H]\right) \right]$, $\alpha\in [0,1]$,
which is a one-parameter extension of the Wigner-Yanase skew information $I_{\rho}(H)$,
 with respect to $\rho$ was successfully proven by E.H.Lieb in \cite{Lie}.  We have the relation between $I_{\rho}(H)$ and $V_{\rho}(H)$ such that
$0\leq I_{\rho}(H) \leq V_{\rho}(H)$ so that it is quite natural to consider that we have the further sharpened uncertainty
 relation for the Wigner-Yanase skew information:
$$I_{\rho}(A)I_{\rho}(B) \geq \frac{1}{4}\vert Tr[\rho[A,B]]\vert^2.$$
However, the above relation failed. (See \cite{Luo0,Kos,YFK}.)
Then, S.Luo introduced the quantity $U_{\rho}(H)$ representing a quantum uncertainty excluding the classical mixture:
\begin{equation}  \label{cla_mix}
U_{\rho}(H)  \equiv \sqrt{V_{\rho}(H)^2 -\left( V_{\rho}(H)-I_{\rho}(H)\right)^2},
\end{equation}
then he succeeded to show a new Heisenberg uncertainty relation on $U_{\rho}(H)$ in \cite{Luo1}:
\begin{equation}   \label{UL_U}
U_{\rho}(A)U_{\rho}(B) \geq  \frac{1}{4}\vert Tr[\rho[A,B]]\vert^2.
\end{equation}
As stated in \cite{Luo1}, the physical meaning of the quantity $U_{\rho}(H)$ can be interpreted as follows.
For a mixed state $\rho$, the variance $V_{\rho}(H)$ has both classical mixture and quantum uncertainty.
Also, the Wigner-Yanase skew information $I_{\rho}(H)$ represents a kind of quantum uncertainty \cite{Luo2,Luo3}. Thus, the difference
$V_{\rho}(H) - I_{\rho}(H)$ has a classical mixture so that we can regard that the quantity $U_{\rho}(H)$  
has a quantum uncertainty excluding a classical mixture. Therefore it is meaningful and suitable to study an uncertainty
relation for a mixed state, by the use of  the quantity $U_{\rho}(H)$.

Recently, K.Yanagi gave a one-parameter extension of the inequality (\ref{UL_U}) in \cite{Yanagi}, 
using the Wigner-Yanase-Dyson skew information $I_{\rho,\alpha}(H)$.
Note that we have the following ordering  among three quantities: 
\begin{equation}  \label{note1}
0 \leq I_{\rho}(H)\leq U_{\rho}(H)\leq V_{\rho}(H).
\end{equation}
The inequality (\ref{UL_U}) is a refinement of the original Heisenberg's uncertainty relation (\ref{HUL}) in the sense of the above ordering (\ref{note1}).

In this brief report, we show the further strong inequality (Schr\"odinger type uncertainty relation) 
for the quantity $U_{\rho}(H)$ representing a quantum uncertainty.

\section{Main results} \label{sec2}
To show our main theorem, we prepare the definition for a few quantities and a lemma representing properties on their quantities.
\begin{Def} \label{def1}
For a quantum state $\rho$ and an observable $H$, we define the following quantities.
\begin{itemize}
\item[(i)] The Wigner-Yanase skew information:
$$
I_{\rho}(H) \equiv \frac{1}{2} Tr\left[(i[\rho^{1/2},H_0])^2\right] = Tr[\rho H_0^2] -Tr[\rho^{1/2} H_0 \rho^{1/2} H_0],
$$
where $H_0\equiv H-Tr[\rho H] I$ and $[X,Y] \equiv XY-YX$ is a commutator.
\item[(ii)]  The quantity associated to the Wigner-Yanase skew information:
$$
J_{\rho}(H) \equiv \frac{1}{2} Tr\left[\left(\left\{\rho^{1/2},H_0\right\}\right)^2\right] = Tr[\rho H_0^2] +Tr[\rho^{1/2} H_0 \rho^{1/2} H_0],
$$
where $\left\{X,Y\right\} \equiv XY+YX$ is an anti-commutator.
\item[(iii)] The quantity representing a quantum uncertainty:
$$
U_{\rho}(H) \equiv \sqrt{V_{\rho}(H)^2-(V_{\rho}(H)-I_{\rho}(H))^2}.
$$
\end{itemize}
\end{Def}

For two quantities $I_{\rho}(H)$ and $J_{\rho}(H)$, by simple calculations, we have
$$
I_{\rho}(H) = Tr[\rho H^2] -Tr[\rho^{1/2} H \rho^{1/2} H]
$$
and
\begin{eqnarray}
J_{\rho}(H) &=& Tr[\rho H^2] + Tr[\rho^{1/2} H \rho^{1/2} H] -2(Tr[\rho H])^2 \nonumber \\
&=& 2 V_{\rho}(H) - I_{\rho}(H),   \label{J_exp}
\end{eqnarray}
which implies $I_{\rho}(H) \leq J_{\rho} (H)$.
In addition, we have the following relations.
\begin{Lem} \label{lem}
\begin{itemize}
\item[(i)]  For a quantum state $\rho$ and an observable $H$, we have the following relation among $I_{\rho}(H)$, $J_{\rho}(H)$ and $U_{\rho}(H)$:
$$
U_{\rho}(H) = \sqrt{I_{\rho}(H)J_{\rho}(H) }.
$$
\item[(ii)] For a spectral decomposition of $\rho = \sum_{j=1}^{\infty} \lambda_j \vert \phi_j \rangle \langle \phi_j \vert$, 
putting $h_{ij} \equiv \langle \phi_i \vert H_0 \vert \phi_j \rangle$, we have
$$
I_{\rho}(H) = \sum_{i<j} \left(\sqrt{\lambda_i}-\sqrt{\lambda_j} \right)^2 \vert h_{ij} \vert^2, 
$$
\item[(iii)] For a spectral decomposition of $\rho = \sum_{j=1}^{\infty} \lambda_j \vert \phi_j \rangle \langle \phi_j \vert$, 
putting $h_{ij} \equiv \langle \phi_i \vert H_0 \vert \phi_j \rangle$, we have
$$
J_{\rho}(H) \geq  \sum_{i<j} \left(\sqrt{\lambda_i}+\sqrt{\lambda_j} \right)^2 \vert h_{ij} \vert^2. 
$$
\end{itemize}
\end{Lem}

(i) immediately follows from Eq.(\ref{J_exp}). 
See \cite{Yanagi} for the proofs of (ii) and (iii). 

\begin{The} \label{the}
For a quantum state (density operator) $\rho$ and two observables (self-adjoint operators) $A$ and $B$, we have
\begin{equation} \label{conjecture}
U_\rho(A)U_\rho(B)-|Re\left\{ Corr_{\rho}(A,B)\right\}|^2 \geq  \frac{1}{4}|Tr[\rho[A,B]]|^2,
\end{equation}
where the correlation measure is defined by
$$
Corr_{\rho}(X,Y) \equiv Tr[\rho X^*Y] -Tr[\rho^{1/2} X^*\rho^{1/2}Y]
$$
for any operators $X$ and $Y$. 
\end{The}

{\it Proof}:
We take a spectral decomposition $\rho = \sum_{j=1}^{\infty} \lambda_j \vert \phi_j \rangle \langle \phi_j \vert$.
If we put $a_{ij}=  \langle \phi_i\vert A_0 \vert \phi_j \rangle$ and $b_{ji}= \langle \phi_j\vert B_0 \vert \phi_i \rangle$,
where $A_0=A-Tr[\rho A] I$ and $B_0=B-Tr[\rho B] I$, then we have
\begin{eqnarray*}
Corr_{\rho}(A,B) &=& Tr[\rho A B] -Tr[\rho^{1/2} A \rho^{1/2} B] \\
&=& Tr[\rho A_0 B_0] -Tr[\rho^{1/2} A_0 \rho^{1/2} B_0] \\
&=& \sum_{i,j=1}^{\infty}(\lambda_i-\lambda_i^{1/2}\lambda_j^{1/2}) a_{ij} b_{ji}\\
&=& \sum_{i\neq j} (\lambda_i-\lambda_i^{1/2}\lambda_j^{1/2}) a_{ij} b_{ji}\\
& =& \sum_{i < j}  \left\{ (\lambda_i-\lambda_i^{1/2}\lambda_j^{1/2}) a_{ij} b_{ji} + (\lambda_j-\lambda_j^{1/2}\lambda_i^{1/2}) a_{ji} b_{ij} \right\}.
\end{eqnarray*}
Thus we have
$$
\vert Corr_{\rho}(A,B)\vert  \leq \sum_{i < j} \left\{  \vert \lambda_i -\lambda_i^{1/2} \lambda_j^{1/2} \vert \vert a_{ij} \vert \vert b_{ji} \vert
+  \vert \lambda_j -\lambda_j^{1/2} \lambda_i^{1/2} \vert \vert a_{ji} \vert \vert b_{ij} \vert \right\}.
$$
Since $\vert a_{ij} \vert = \vert a_{ji} \vert$ and $\vert b_{ij} \vert = \vert b_{ji} \vert$, 
taking a square of both sides and then using Schwarz inequality and  Lemma \ref{lem}, we have
\begin{eqnarray*}
\vert Corr_{\rho}(A,B)\vert ^2 &\leq & \left\{ \sum_{i < j}  \left\{  \vert \lambda_i -\lambda_i^{1/2} \lambda_j^{1/2} \vert 
+  \vert \lambda_j -\lambda_j^{1/2} \lambda_i^{1/2} \vert \right\} \vert a_{ij} \vert \vert b_{ji} \vert \right\}^2 \\
&=& \left\{ \sum_{i<j} \left( \lambda_i^{1/2}+\lambda_{j}^{1/2} \right) \vert \lambda_i^{1/2} -\lambda_j^{1/2} \vert \vert a_{ij} \vert \vert b_{ji} \vert \right\}^2 \\
& \leq & \left\{\sum_{i<j} \left( \sqrt{\lambda_i} - \sqrt{\lambda_j }\right)^2 \vert a_{ij} \vert^2\right\} 
\left\{\sum_{i<j} \left( \sqrt{\lambda_i} + \sqrt{\lambda_j }\right)^2 \vert b_{ij} \vert^2\right\} \\
&\leq & I_{\rho}(A) J_{\rho}(B)
\end{eqnarray*}
By the similar way, we also have
$$
\vert Corr_{\rho}(A,B)\vert ^2 \leq I_{\rho}(B) J_{\rho}(A)
$$
Thus we have
$$
\vert Corr_{\rho}(A,B)\vert ^2 \leq U_{\rho}(A) U_{\rho}(B),
$$
which is equivalent to the inequality:
$$
U_\rho(A)U_\rho(B)-|Re\left\{ Corr_{\rho}(A,B)\right\}|^2 \geq  \frac{1}{4}|Tr[\rho[A,B]]|^2,
$$
since we have
$$
\vert Im\left\{Corr_{\rho}(A,B) \right\} \vert^2 = \frac{1}{4} \vert Tr[\rho [A,B]] \vert ^2.
$$

\hfill \qed

Theorem \ref{the} improves the uncertainty relation (\ref{UL_U}) shown in \cite{Luo1}, in the sense that the upper bound of the right hand side
of our inequality (\ref{conjecture}) is tighter than that of S.Luo's one (\ref{UL_U}). 


\begin{Rem} \label{rem01}
For a pure state $\rho = \vert \varphi \rangle \langle \varphi \vert$,
we have $I_{\rho}(H) = V_{\rho}(H)$ which implies
$U_{\rho}(H) = V_{\rho}(H)$ for an observable $H$ and
$Corr_{\rho}(A,B) = Cov_{\rho}(A,B)$ for two observables $A$ and $B$.
Therefore our Theorem \ref{the} coincides with the Schr\"odinger uncertainty relation (\ref{S_UL})
 for a  particular case that a given quantum state is a pure state, 
$\rho = \vert \varphi \rangle \langle \varphi \vert$.
\end{Rem}

\begin{Rem} \label{rem02}
As a similar problem, we may consider the following uncertainty relation:
$$
U_\rho(A)U_\rho(B)-|Re\left\{ Cov_\rho(A,B)\right\}|^2 \geq  \frac{1}{4}|Tr[\rho[A,B]]|^2.
$$
However, the above inequality does not hold in general, since we have a counter-example as follows.
We take
\[
\rho  = \frac{1}{4}\left( {\begin{array}{*{20}c}
   1 & 0  \\
   0 & 3  \\
\end{array}} \right),
A = \left( {\begin{array}{*{20}c}
   2 & 1  \\
   1 & 2  \\
\end{array}} \right),B = \left( {\begin{array}{*{20}c}
   0 & 1  \\
   1 & 0  \\
\end{array}} \right),
\]
then we have
$$
U_\rho(A)U_\rho(B)-|Re\left\{ Cov_{\rho}(A,B)\right\}|^2 - \frac{1}{4}|Tr[\rho[A,B]]|^2 = -\frac{3}{4}.
$$
\end{Rem}

\begin{Rem} \label{rem03}
From Theorem \ref{the} and Remark \ref{rem02}, we may expect that the following inequality holds:
\begin{equation}
\vert Re \left\{ Cov_{\rho}(A,B) \right\}\vert^2 \geq \vert Re \left\{ Corr_{\rho}(A,B) \right\}\vert^2. 
\end{equation} 
However, the above inequality does not hold in general, since we have a counter-example as follows.
We take
\[
\rho  = \frac{1}{{10}}\left( {\begin{array}{*{20}c}
   5 & 4  \\
   4 & 5  \\
\end{array}} \right),A = \left( {\begin{array}{*{20}c}
   4 & 4  \\
   4 & 1  \\
\end{array}} \right),B = \left( {\begin{array}{*{20}c}
   5 & { - 1}  \\
   { - 1} & 2  \\
\end{array}} \right),
\]
then we have
$$
\vert Re \left\{ Cov_{\rho}(A,B) \right\}\vert^2 - \vert Re \left\{ Corr_{\rho}(A,B) \right\}\vert^2 \simeq -0.1539.
$$
Actually, from Theorem \ref{the}, the example in Remark \ref{rem02} and the above example, 
we find that there is no ordering between $\vert Re \left\{ Cov_{\rho}(A,B) \right\}\vert^2$ and $\vert Re \left\{ Corr_{\rho}(A,B) \right\}\vert^2 $.
\end{Rem}

\begin{Rem} \label{lem04}
The example given in Remark \ref{rem02} shows
$$ V_\rho(A)V_\rho(B)-|Re\left\{ Cov_{\rho}(A,B)\right\}|^2  -\left( U_\rho(A)U_\rho(B)-|Re\left\{ Corr_{\rho}(A,B)\right\}|^2  \right) \simeq -0.232051.$$
The example given in Remark \ref{rem03} also shows
$$V_\rho(A)V_\rho(B)-|Re\left\{ Cov_{\rho}(A,B)\right\}|^2  -\left(U_\rho(A)U_\rho(B)-|Re\left\{ Corr_{\rho}(A,B)\right\}|^2  \right) \simeq 13.7862.$$
Therefore there is no ordering between $V_\rho(A)V_\rho(B)-|Re\left\{ Cov_{\rho}(A,B)\right\}|^2  $ and $U_\rho(A)U_\rho(B)-|Re\left\{ Corr_{\rho}(A,B)\right\}|^2  $
so that we can conclude that neither the inequality (\ref{S_UL}) nor the inequality (\ref{conjecture}) is uniformly better than the other.
\end{Rem}

\section{Conclusion}
As we have seen, we proved a new Schr\"odinger type uncertainty relation for a quantum state (generally a mixed state).
Our result coincides with the original  Schr\"odinger uncertainty relation for a particular case that a quantum state is a pure state.
In addition, our result improves the uncertainty relation shown in \cite{Luo1} and obviously does the original Heisenberg uncertainty relation.
Moreover,  it is impossible to conclude that our result is better than the original  Schr\"odinger uncertainty relation for a mixed state, in the sense of
finding a tighter upper bound for $\frac{1}{4} \vert Tr\left[\rho [A,B]\right]\vert^2$, where $Tr[\rho [A,B]]$ can be regarded as an average of
the commutator $[A,B]$ for two observables $A$ and $B$ in a quantum state $\rho$. 
However, in other words, it is also impossible to conclude that our result is a trivial one, 
since there is no ordering between the left hand side of the inequality (\ref{S_UL}) and that of one (\ref{conjecture}).

\section*{Acknowledgement}
The author was supported in part by the Japanese Ministry of Education, Science, Sports and Culture, Grant-in-Aid for
Encouragement of Young Scientists (B), 20740067

\end{document}